\documentclass[12pt, a4paper]{article}
\usepackage{lineno}
\usepackage{color}
\usepackage{graphicx}
\usepackage[toc,page]{appendix}
\usepackage{caption}
\usepackage{amsmath}
\usepackage{subcaption}
\usepackage{hyperref}
\usepackage{mathtools}
\usepackage{float}
\usepackage{rotating}
\usepackage{url}
\usepackage{enumitem}
\usepackage[table]{xcolor}
\usepackage{multirow}
\usepackage[font=scriptsize,labelfont=bf,skip=6pt]{caption}
\usepackage[T1]{fontenc}
\usepackage{authblk}
\font\myfont=cmr12 at 15pt

\begin{document}
\title{\myfont{Performance of prototype GE1$\slash$1 chambers for the CMS muon spectrometer upgrade}}

\author[12]{M. Abbas} 
\author[18]{M. Abbrescia}
\author[9]{H. Abdalla}
\author[9]{S. Abu Zeid}
\author[5]{A. Agapitos}
\author[27]{A. Ahmad}
\author[9]{A. Ahmed}
\author[15]{A. Ahmed}
\author[27]{W. Ahmed}
\author[27]{I. Asghar}

% ...
\affil[5]{Peking University, Beijing, China} %
% ...
\affil[9]{Academy of Scientific Research and Technology - ENHEP, Cairo, Egypt} %
% ...
\affil[12]{Karlsruhe Institute of Technology, Karlsruhe, Germany}
% ...
\affil[15]{Delhi University, Delhi, India}
% ...
\affil[18]{Politecnico di Bari, Universit\`{a} di Bari and INFN Sezione di Bari, Bari, Italy}%
% ...
\affil[27]{National Center for Physics, Islamabad, Pakistan}
% ...

\maketitle
\begin{abstract}

The high-luminosity phase of the Large Hadron Collider (HL-LHC) will result in ten times higher particle background  than measured during the first phase of LHC operation. In order to fully exploit the highly-demanding operating conditions during HL-LHC, the Compact Muon Solenoid (CMS) Collaboration will use Gas Electron Multiplier (GEM) detector technology. The technology will be integrated into the innermost region of the forward muon spectrometer of CMS as an additional muon station called GE1$\slash$1. The primary purpose of this auxiliary station is to help in muon reconstruction and to control level-1 muon trigger rates in the pseudo-rapidity region $1.6 <\left|\eta\right|<2.2$. The new station will contain trapezoidal-shaped GEM detectors called GE1$\slash$1 chambers. The design of these  chambers is finalized, and the installation is in progress during the Long Shutdown phase two (LS-2) that started in 2019.  Several full-size prototypes were built and operated successfully in various test beams at CERN. We describe performance measurements such as gain, efficiency, and time resolution of these prototype chambers, developed after years of R$\&$D, and summarize their behavior in different gas compositions as a function of the applied voltage. \\ 
\end{abstract}
\clearpage 

\tableofcontents

\clearpage 
\section{Introduction}

\noindent The High Luminosity upgrade of the LHC (HL-LHC) will provide
p-p collisions with 
center-of-mass energy of 14 TeV and instantaneous luminosity ($\mathcal{L}$) up to or above 5 $\times$ 10$^{34}$ cm$^{-2}$s$^{-1}$. The increase in the collision rate will affect the operational conditions in HL-LHC due to the increase in pileup and radiation background. It will also pose a challenge to maintain an efficient and reliable trigger particularly in the region $\left|\eta\right|>1.6$. The high-radiation background may accelerate aging of the current muon system and may cause performance losses, dead regions and degradation of the efficiency of online event selection due to bandwidth limitations.

\begin{figure}[!ht]
    \centering
          \includegraphics[width=9cm, height=6.2cm]{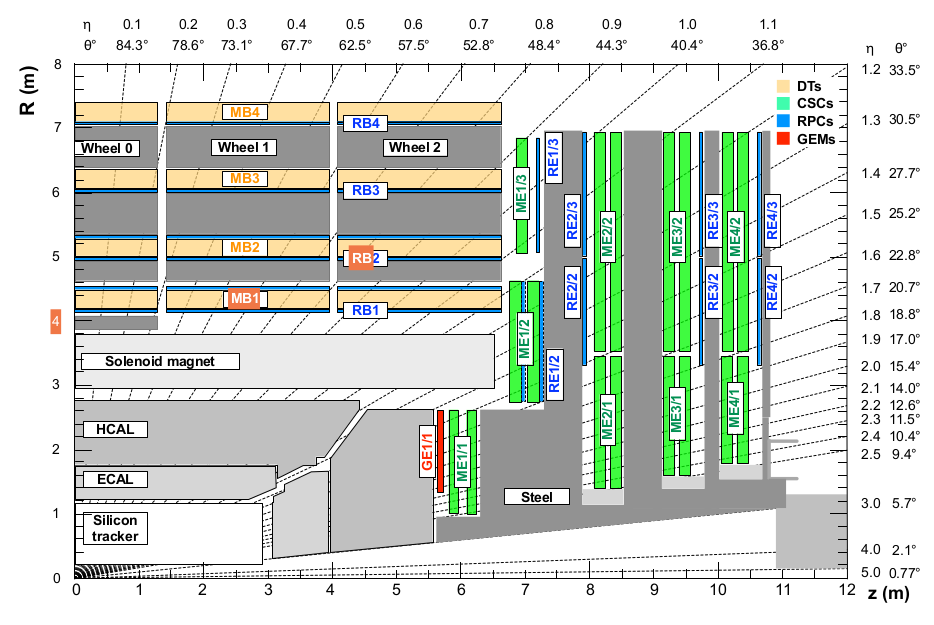} 
   \caption{Sliced view of CMS detector highlighting (in Red) the location of the GE1$\slash$1 in the pseudo-rapidity region $1.6 <\left|\eta\right|<2.1$~\cite{cmsTdr}. } \label{fig:cms}
\end{figure}

\noindent The CMS Collaboration is preparing for the upgrade of the
current muon system scheduled in 2019 to perpetuate its high level of
performance. A quadrant of the CMS muon system 
with existing
detectors and proposed extensions is shown in Figure~\ref{fig:cms}. In
the $1.6 <\left|\eta\right|<2.4$ forward end-cap region, currently
only Cathode Strip Chambers (CSC) are installed. To enhance muon
trigger and reconstruction capabilities, 
large-area GEM detectors \cite{one01, cmsTdr} will be installed in
this region. These detectors play a significant role in the
instrumentation of particle physics experiments and are known to have
high performance with spatial resolution better than 70 $\mu$m, rate
capability 
of order MHz/cm$^2$, 
and high tolerance to radiation in strong radiation background
environments. The integration of these new detectors together with the
existing CSC system will highly improve the muon trigger momentum
resolution due to an increase in the lever arm for the measurement of
the muon bending angle. In particular, the new station to be installed
is GE1$\slash$1\footnote{In ``GE1$\slash$1'', the ``G'' stands for GEM and the ``E'' for  Endcap; the first ``1'' corresponds to the first muon station and the second ``1'' to the first, innermost ring of the station.}, which would be equipped with a specific type of GEM detectors named as GE1/1 chambers.

\noindent We  present performance studies such as of gain, efficiency, time resolution, and discharge probability of GEM GE1$\slash$1 chambers and further describe their behavior for standard CMS operating conditions.  The document is structured as follows: the first two sections describe the preliminary details such as the design of GE1$\slash$1 chambers and the CMS GEM upgrade. The third to seventh sections describe the performance studies, which are followed by the summary in which the recommended operating conditions of the GE1/1 chambers for CMS are provided. 

\begin{figure}[!ht]
    \centering
           \includegraphics[width=9cm, height=4.2cm] {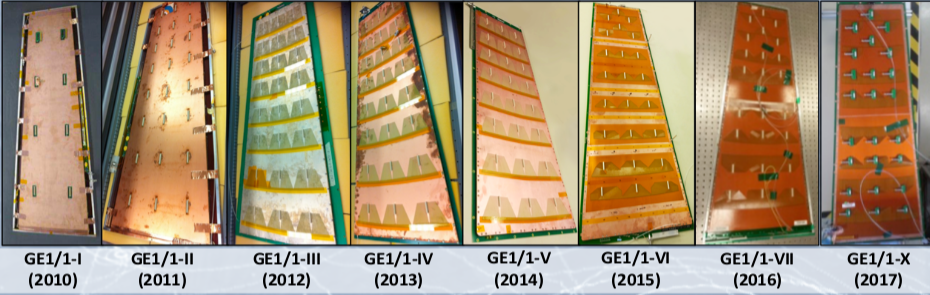}
   \caption{Evolution of GE1$\slash$1 detector since 2010~\cite{cmsTdr} from generation-I to generation-X (2018).} \label{fig:evolution}
\end{figure}

Succesive versions of the GE1$\slash$1 chambers have been built by improving their design 
%
%
in each release. 
Figure~\ref{fig:evolution} shows the evolution of the 
GE1$\slash$1 detectors since 2010, 
when the
CMS Collaboration proposed their use in 
the
muon end-cap region of 
the
CMS detector.
The latest version is generation-X whose design is discussed in Section~\ref{Sec:Design}. The mechanical constraints in the GE1$\slash$1 station 
require
two trapezoidal types of detectors to be used to 
obtain 
maximum detection coverage. 
Long chambers GE1$\slash$1-L 
have a  length of 128.5 cm and 
short chambers GE1$\slash$1-S 
have a length of 113.5 cm. The technical details 
of Short and Long versions, their construction and layout, can be found in~\cite{GEMlayout}. 
Two identical GE1$\slash$1 detectors are combined to form a ``super-chamber'' to obtain two detection planes and thus maximize the detection efficiency and the redundancy of the GE1$\slash$1 layer.

\section{Impact of GE1$\slash$1 upgrade on CMS} \label{Sect:ImpactCMS}

\noindent The introduction of the new station known as GE1$\slash$1 will cover the pseudo-rapidity region  $1.6 <\left|\eta\right|<2.2$ of 
CMS~\cite{cmsTdr} and 
complement the current CSC system. These new chambers are based on GEM technology and can operate at very high rates with good performance. The GE1$\slash$1 station will extend the path length and will provide additional hits that will help to refine the stub reconstruction and improve the momentum resolution. With the new station installed, muon direction will be measured using hit positions in the adjacent GEM GE1$\slash$1 and CSC ME1$\slash$1 chambers. The good position resolution of both the detectors and an increased lever arm formed by the two detectors will allow excellent directional measurement.

\noindent FLUKA simulation studies at  $\mathcal{L}$ =  5 $\times$ 10$^{34}$ cm$^{-2}$s$^{-1}$ are used to assess  the capability of this new technology to cope with background fluxes expected in the high $\eta$ region~\cite{cmsTdr}. 
The background rate is estimated from standalone Geant4 simulation, convoluting the fluxes mentioned above with the chamber sensitivities to background.
The resulting rate is found to be 
of order 1 kHz$\slash$cm$^{2}$, 
orders of magnitude below the rate capability limit of the chambers,
whose gain is stable up to a few MHz$\slash$cm$^{2}$.

\begin{figure}[!ht]
\centering
\includegraphics[width=8.5cm, height=6.5cm]{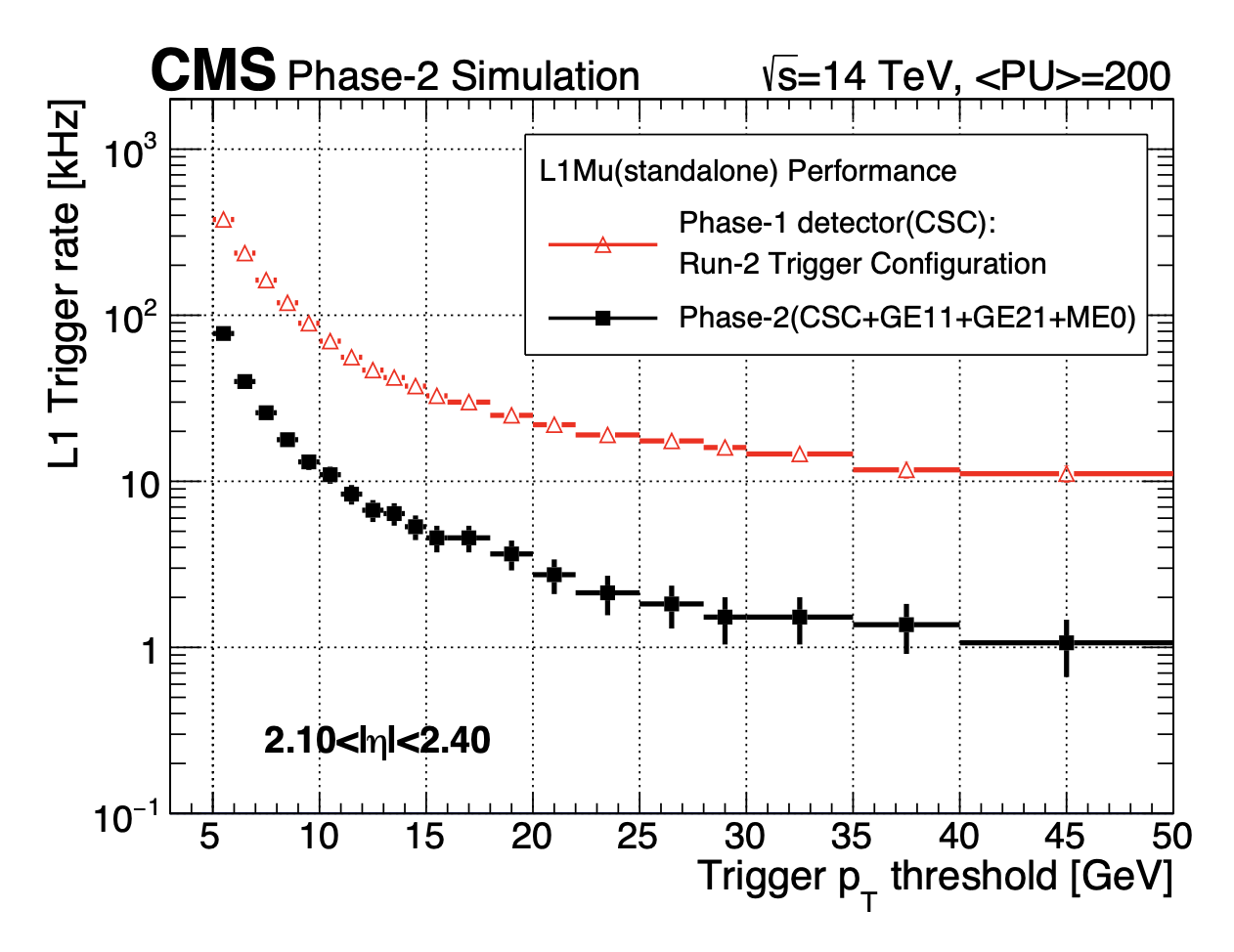}
\caption{Muon trigger rate at Level 1 with and without GE1/1 GEM chambers and additional GEM chambers GE2/1 and ME0 described in reference~\cite{cmsTdr02}.} \label{fig:PhaseIISim}
\end{figure}

\noindent 
Introducing 
the
GE1$\slash$1 muon station 
will help
to reduce
the level-1 (L1) muon trigger rates as shown in Figure~\ref{fig:PhaseIISim}. 
The distance between adjacent CSCs and GEMs will allow determination of the muon $p_{T}$ by measuring the bending angle due to the magnetic field in the first muon station alone. 
This $p_{T}$ measurement, independent from the one based on the muon bending through the 
rest of the 
detector, will allow the maintenance of a low momentum threshold. It 
will
be crucial for a broad spectrum of physics processes whose signatures are characterized by the presence of low $p_{T}$ muons in the final state. 
A few examples include the search for the lepton flavor violating decay $\tau \rightarrow 3 \mu$, Higgs boson decays $h \rightarrow ZZ \rightarrow 4 \mu$ and $h \rightarrow 4 \mu$, $B$ meson decay $B \rightarrow 2 \mu$, and two-Higgs doublet model extended with a scalar singlet (2HDM + S) (Higgs) decay $h \rightarrow aa \rightarrow bb \mu\mu$~\cite{cmsTdr}.

\section{GE1$\slash$1 detector design} \label{Sec:Design}

\noindent A GEM~\cite{one01} is a 50 $\mu$m thick copper-clad polymer (Kapton or Apical NP) foil chemically perforated by a high density of microscopic holes. The copper cladding is on both sides of the foil with a thickness of 5 $\mu$m. The holes in the foil are pierced with double cones with outer diameter 70 $\mu$m, inner diameter 50 $\mu$m, and pitch 140 $\mu$m. Each hole acts as a signal amplifier. Three foils are cascaded to form a detector known as triple GEM to obtain a measurable signal.

\begin{figure}[!ht]
\centering
\includegraphics[width=12.0cm, height=7.5cm ]{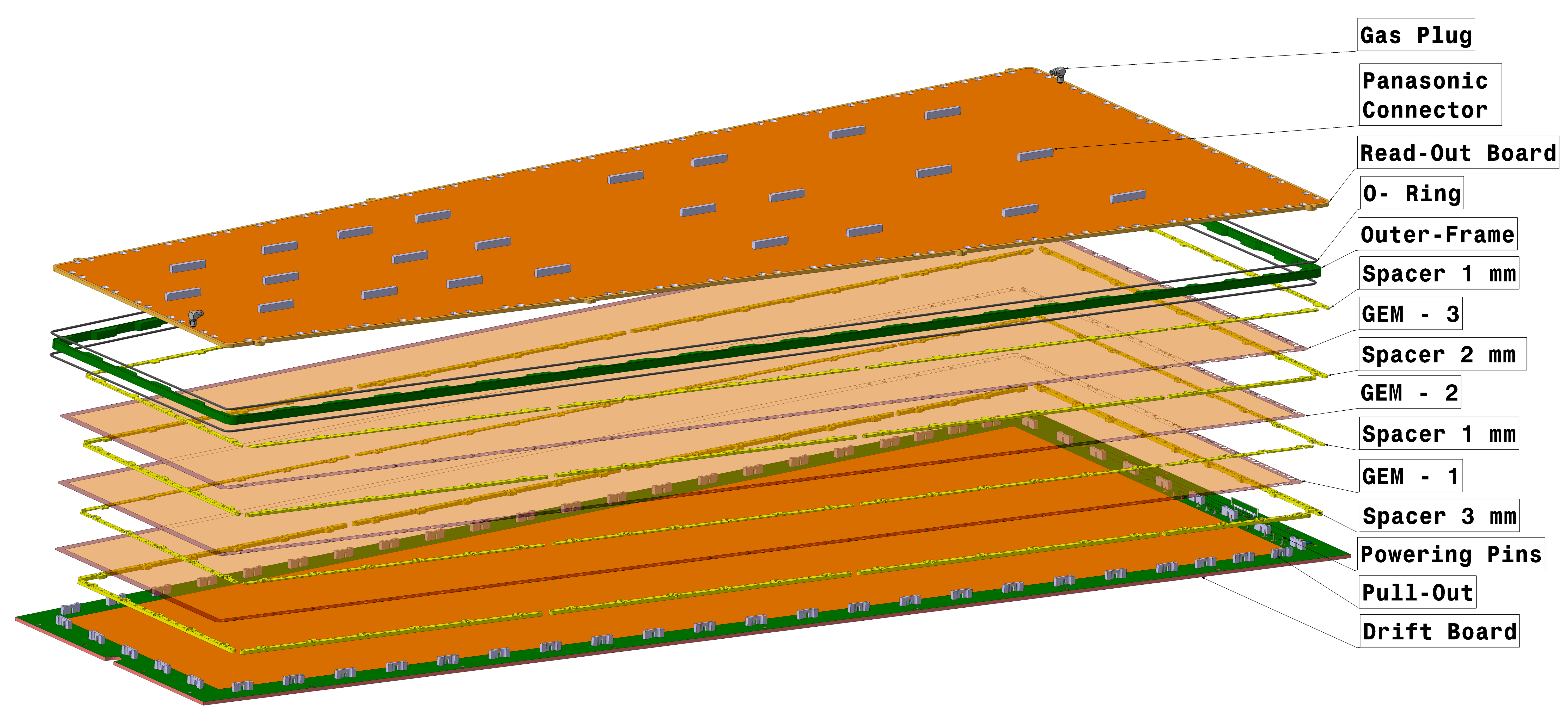}
\includegraphics[width=6.2cm, height=5cm ]{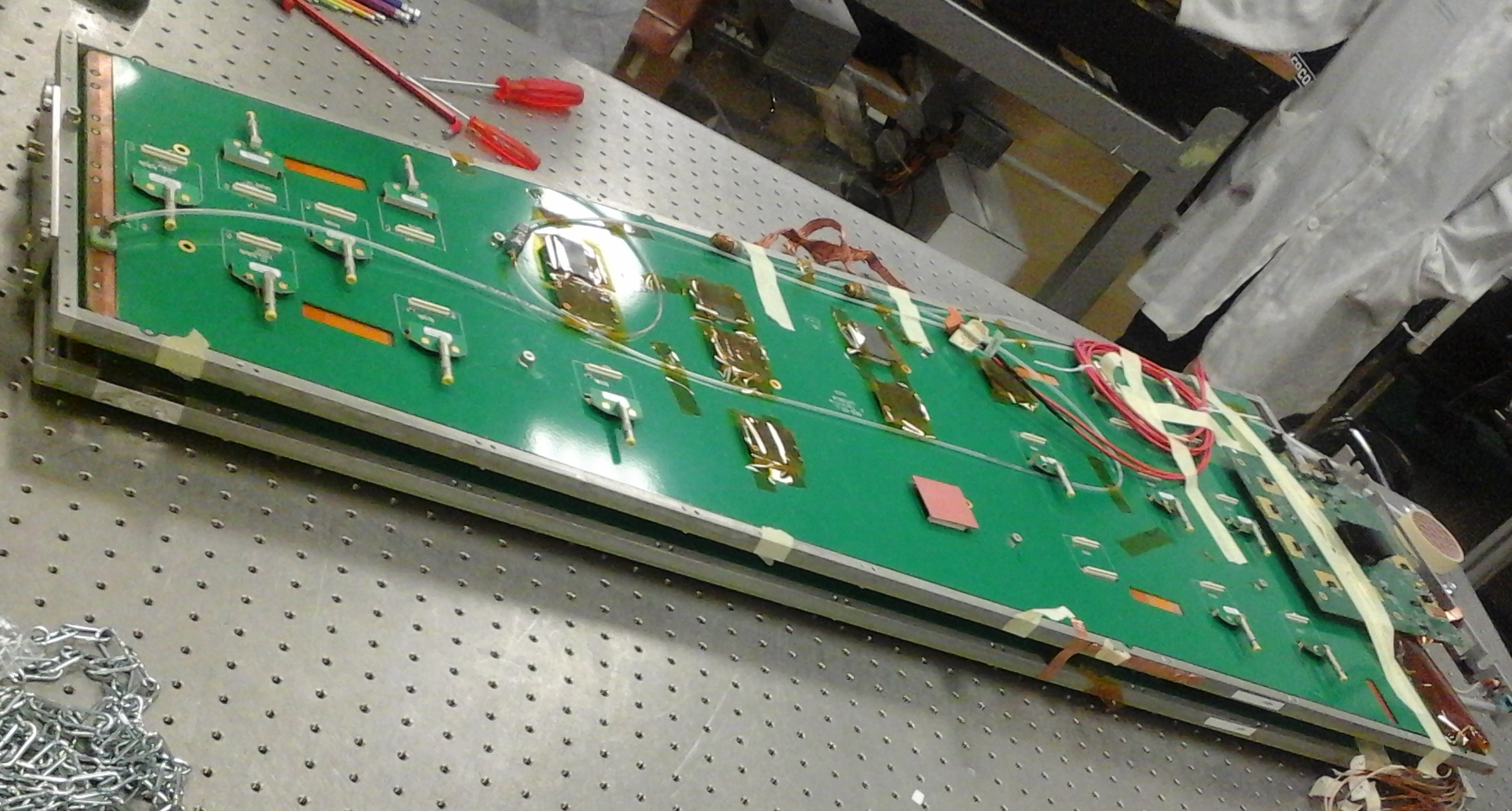}
\includegraphics[width=3.2cm, height=6cm]{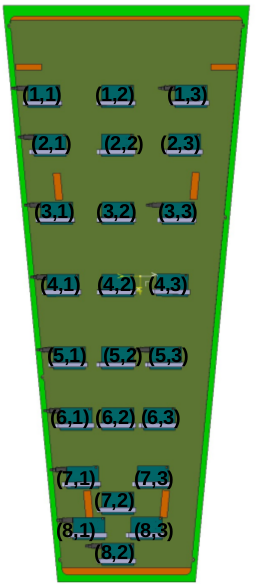}

\caption{(top) GE1$\slash$1 layout and final design. The main components from bottom: drift board mounted all around with stainless steel pull-outs used for stretching of GEM foils, 3 mm frame, first foil, 1 mm frame, second foil, 2 mm frame, third foil, 1 mm frame, first O-ring, external frame, second O-ring and the readout board, (bottom-left) two detectors connected back to back to form a GE1/1 `super-chamber', and (bottom-right) map of the readout board showing 24 ($\eta$, $\phi$) sectors of GE1$\slash$1 chambers.} \label{fig:gem_detector}
\end{figure}

\noindent The construction of the GE1/1 detectors is shown in
Figure~\ref{fig:gem_detector}. The large area GEM foils needed for CMS
are produced at the CERN Micro-Pattern Techniques (MPT) workshop
using a single-mask production technique~\cite{one011}. The surfaces
of GEM foils oriented towards the readout board are a single
continuous conductor whereas the surfaces facing towards the drift
board are segmented into sectors each of area about 100 cm$^{2}$.
This segmentation limits the energy of discharges, helping to prevent
damage that might be large enough to generate a short.
If a short does occur, segmentation reduces the dead area, allowing
the rest of the detector to remain sensitive, and 
reducing the dead time necessary to recharge the affected segment.
More details are given in Section~\ref{Sect:DischargeProbability}.

\noindent The drift board is a trapezoidal-shaped printed circuit board (PCB) serving as drift electrode. The board has an active area coated with a copper layer and lies within the active gas volume. The readout board is also a trapezoidal-shaped PCB with the inner side of the board featuring 3072 trapezoidal readout strips oriented radially along the longer sides of the detector. All the readout strips are connected through metalized vias to the outer side of the board where traces are routed from the vias to readout pads in 8 $\times$ 3 partitions in ($\eta$, $\phi$) as shown in Figure~\ref{fig:gem_detector}. Each $\eta$-partition has 384 strips comprised of three 128-strip sectors in $\phi$. Drift board, readout board, and external frame define the gas volume with gas tightness ensured by an O-ring placed in the groove of the outer frame~\cite{GEMlayout}.\\

\section{Gain Measurements} \label{Sect:GainMeasuremts}

\subsection{Test Setup}

\begin{figure}[!ht]
    \centering
        \includegraphics[width=8.7cm, height=6cm]{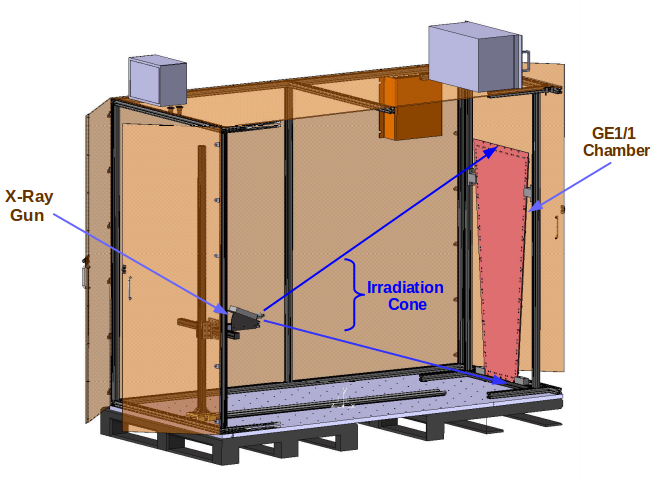}
   \caption{Design of the setup used for gain measurements using an X-ray tube, with a GE1$\slash$1 detector inside the copper chamber. The copper chamber is compeletely closed when the detector is exposed to X-rays.} \label{fig:X_Ray_Station}
\end{figure}

\noindent The detector under test is powered using a programmable high
voltage (HV) power supply (CAEN N1470) that allows a controllable
current limit ($I_{set}$), voltage ramping up and down in steps,
maximum voltage, and trip time (the maximum time the current can
remain over the controllable limit). The power supply delivers a current up to 1 mA with a monitoring resolution of about 50 nA, which allows identification of unusual current fluctuations. The voltages on the foils are provided through a resistive divider network described in~\cite{GEMlayout}.  Measurements are performed by irradiating the detector using a mini AMPTEK X-ray Silver (Ag) target source.  The detector is irradiated within a closed copper chamber, the design of which is shown in Figure~\ref{fig:X_Ray_Station}, which prevents radiation exposure to human beings.

\noindent The chamber is  flushed with a given gas mixture at a flow rate of 5 liters per hour, for at least 5 hours before taking a  measurement.  Two test gas mixtures, Ar$\slash$CO$_{2}$~\cite{GEMOprinciples} and Ar$\slash$CO$_{2}$$\slash$CF$_{4}$~\cite{GEMAdavnces} in proportions of 70$\slash$30 and 45$\slash$15$\slash$40, respectively, are used. The choice of CF$_{4}$ quencher in the latter case is driven by its good timing charateristics~\cite{goodTiming}, non-flammability, and non-corrosivity for metals; it is safer to use than other hydrocarbons such as methane. 

\noindent The effective gas gain is measured by exposing the detector
to an X-ray source with a silver target for generating X-rays. The
incident X-rays consist of silver $K_{\alpha}$ and $K_{\beta}$ peaks
(centered around the energies of 22 and 25 keV) over an electron
bremsstrahlung continuum background. The X-ray photons are absorbed by
the copper atoms of the drift electrode which in turn, emits copper
X-ray photons of 8 keV energy while returning to the ground state. The
X-ray photons emitted by the copper are then converted by the
photoelectric effect in the active gas volume. The resulting spectrum
is thus a convolution of the energies of the incident X-rays photons
interacting in the active gas volume of the detector, the
bremsstrahlung continuum background, and a small fraction of
unconverted silver $K_{\alpha}$ and $K_{\beta}$ lines.  The gain can
be measured in each gas composition by comparing the primary current
I$_{p}$ induced in the drift gap by the X-ray source, and the
amplified output current (I$_{o}$) induced on the readout board.

\noindent
Even though the X-ray photon interaction rate is of order kHz, the primary current is small;
it cannot be measured accurately directly from the drift electrode, which is held at high voltage and prone to noise. Instead, the primary current is determined from the product of the
count rate ($R$), the number of primary electrons per X-ray photon ($n_{e^{-}}$), and the electron charge ($e^{-}$).
The output current from a given sector in ($\eta$, $\phi$) is measured by combining the output from all 128 readout strips in the sector. 
%
%
The combined signal is read out using a charge sensitive pre-amplifier (ORTEC 142PC),  followed, in turn, by an amplifier/shaper unit (ORTEC 474), discriminator (Lecroy 623A), and scaler unit.
The rate plateau is obtained by ramping up the detector HV. 
The output current is measured using a pico-ammeter connected to a Keithley Electrometer Model 6487. The data are recorded with a Labview program via a General Purpose Interface Bus (GPIB). 

\subsection{Results} \label{Sect:Results}

\begin{figure}[!ht]
    \centering  
      \includegraphics[width=8cm, height=6.5cm ]{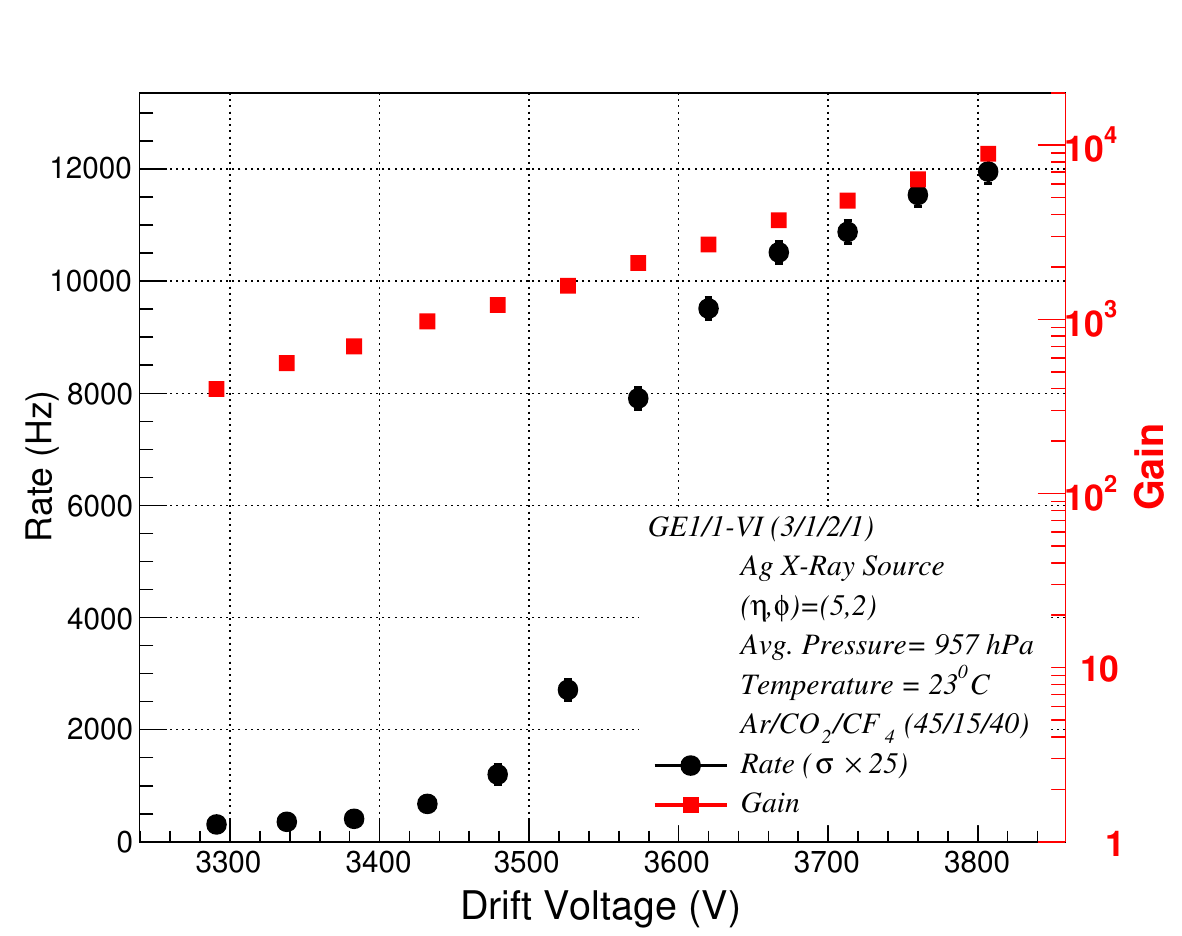} 
   \caption{Gain and rate measurements of sixth generation chamber (GE1$\slash$1-VI) while reading a particular ($\eta$, $\phi$)  = (5, 2) sector. The error bars on the measured rate are Gaussian one sigma uncertainties which are very small and hence are multiplied by a factor of 25 ($\sigma$ $\times$ 25) so as to be visible on the rate curve.} \label{fig:Gain_Rate}
\end{figure}

\noindent An example measurement of rate and gain for the (sixth generation) GE1$\slash$1-VI detector is shown in Figure~\ref{fig:Gain_Rate} for a gas mixture of Ar$\slash$CO$_{2}$$\slash$CF$_{4}$.

\begin{figure}[!ht]
    \centering
        \includegraphics[width=7cm, height=5.4cm ]{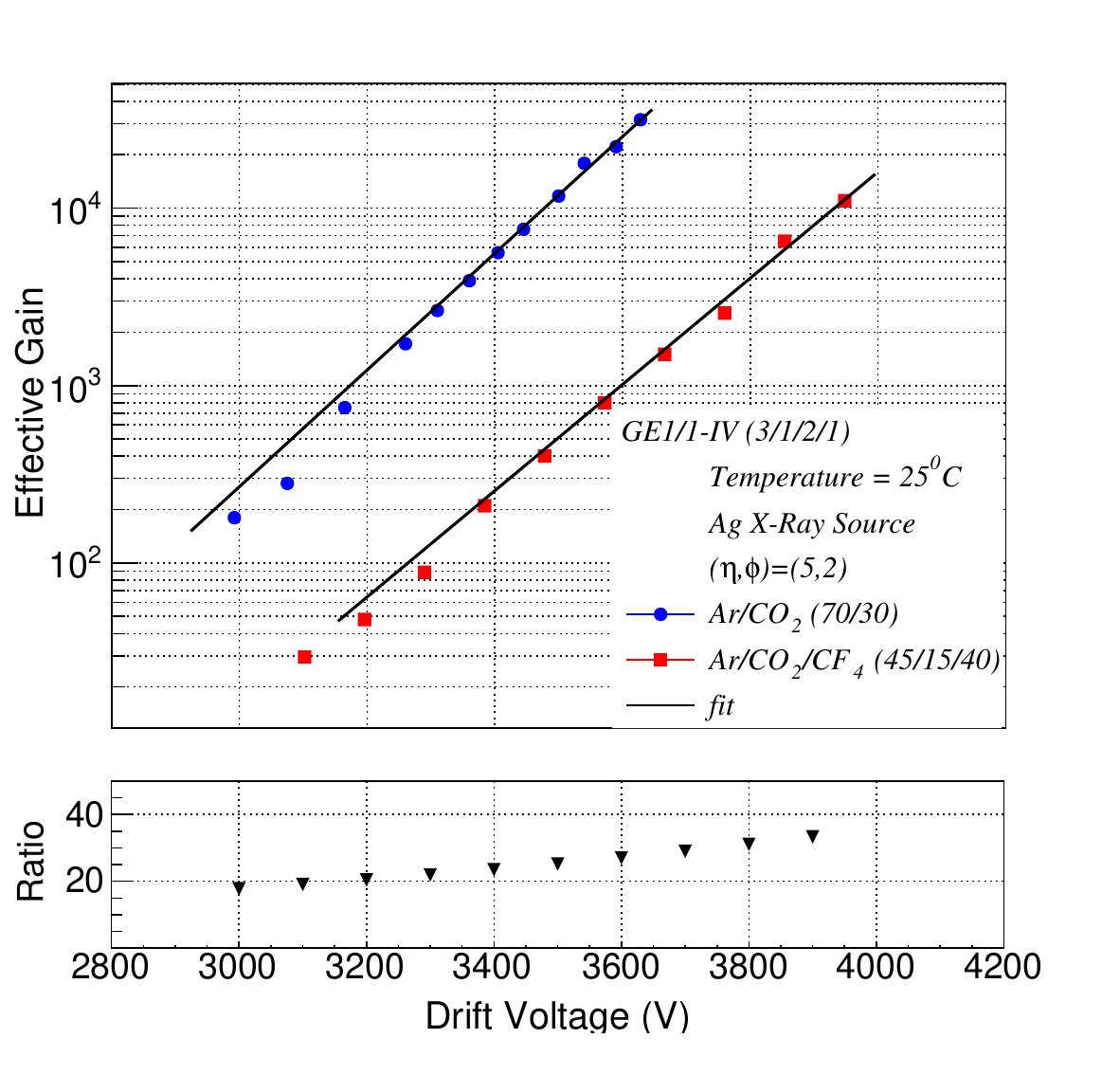} 
        \includegraphics[width=7cm, height=5.4cm]{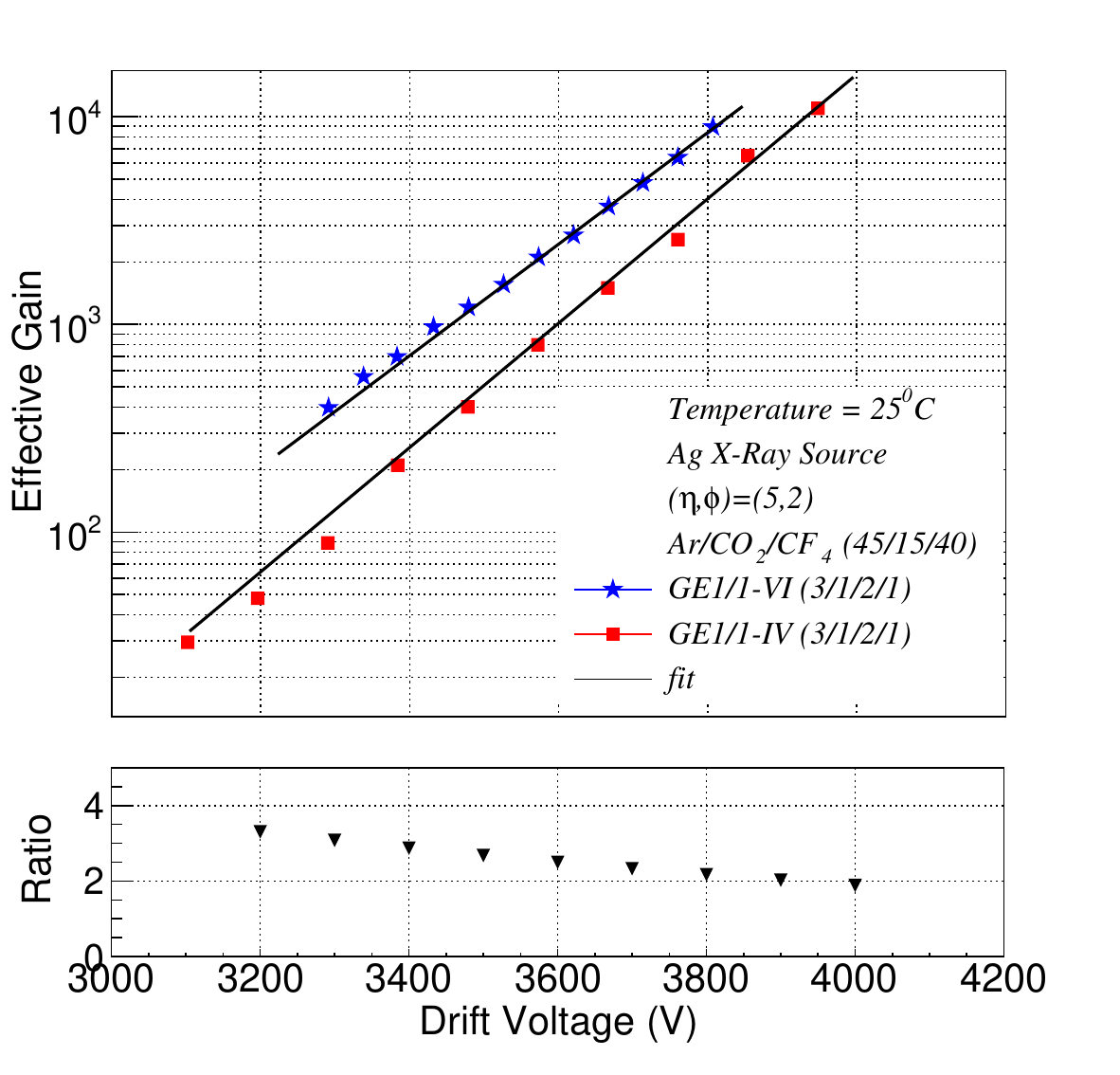}
   \caption{(top) Gain of fourth generation GE1$\slash$1-IV detector for the gas mixtures Ar$\slash$CO$_{2}$ (70$\slash$30) and Ar$\slash$CO$_{2}$$\slash$CF$_{4}$ (45$\slash$15$\slash$40). (bottom) Observed gains of fourth generation GE1$\slash$1-IV and sixth generation GE1$\slash$1-VI detectors for Ar$\slash$CO$_{2}$$\slash$CF$_{4}$ (45$\slash$15$\slash$40). Points represent the data and the solid lines a fit to the observed data. The ratio plots have been calculated by fitting observed data and using the fit equations to interpolate into the regions of missing data points while taking the ratio between the gains corresponding to Ar$\slash$CO$_{2}$ and Ar$\slash$CO$_{2}$$\slash$CF$_{4}$.} \label{fig:Gain_IV}
\end{figure}

\noindent Another example result is shown for a (fourth generation) GE1$\slash$1-IV detector, presented in Figure~\ref{fig:Gain_IV}, for the gas mixtures Ar$\slash$CO$_{2}$  and Ar$\slash$CO$_{2}$$\slash$CF$_{4}$. It is observed that the gain for GE1$\slash$1-IV is higher in Ar$\slash$CO$_{2}$ compared to Ar$\slash$CO$_{2}$$\slash$CF$_{4}$. Mixtures with carbon tetra-fluoride (CF$_{4}$) increase  the electron drift velocity, improving the detector time resolution, but require higher operating voltages because CF$_{4}$ is an electron quencher and its addition to Ar$\slash$CO$_{2}$ reduces the number of electrons participating in the signal formation.  Consequently, at comparable voltages, the gain with Ar$\slash$CO$_{2}$ mixture is higher than that for Ar$\slash$CO$_{2}$$\slash$CF$_{4}$.


\noindent Gain measurements for GE1$\slash$1-IV and GE1$\slash$1-VI detectors 
with gas composition Ar$\slash$CO$_{2}$$\slash$CF$_{4}$ are compared
in Figure \ref{fig:Gain_IV}. The gain is seen to be higher for
GE1$\slash$1-VI than for GE1$\slash$1-IV; the difference is due to the orientation of the GEM foils. Both GE1/1-VI and GE1/1-IV have single-mask GEM foils with holes that are asymmetrically bi-conical in shape. The holes with the  narrow opening facing the incident radiation (chosen for CMS production) show higher gain than when the wider opening of the holes face the incident radiation~\cite{SingleMask01, SingleMask02}.

\noindent
Gain uniformity over the active surface of the GE1/1 detectors is one of the crucial parameters for optimal performance. Non-uniformity can result from factors such as variations in hole diameter or variations  in gas gap width due to improper stretching; the latter is more important for GE1/1 chambers. Therefore, response uniformity  across  the  surface of all the GE1/1 chambers built in 2017 has been measured by recording  the  pulse height  of  every  readout  strip and is observed to be within 15\%~\cite{responseUniformity}. This response uniformity results in efficiency and time resolution that meet operational requirements.

\section{Efficiency and Timing Measurements} \label{Sect:EfficiencyTiming}
%
%
\subsection{Beam Facility}

The prototype GEM detectors were tested in the CERN H4 beam, extracted from the SPS (Super Proton Synchrotron).  A secondary beam consisting of pions and their decay products is produced when the primary proton beam strikes a beryllium target.  The secondary beam is filtered by collimators to produce a beam of muons of energy 150 GeV, which are minimum ionizing particles (MIPs). 



\subsection{Test Setup} \label{Sect:TestSetup}

\begin{figure}[hbtp]
    \centering
	\includegraphics[width=7.5cm, height=4cm]{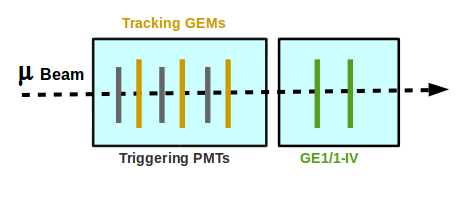}
        \includegraphics[width=4.2cm, height=5.2cm]{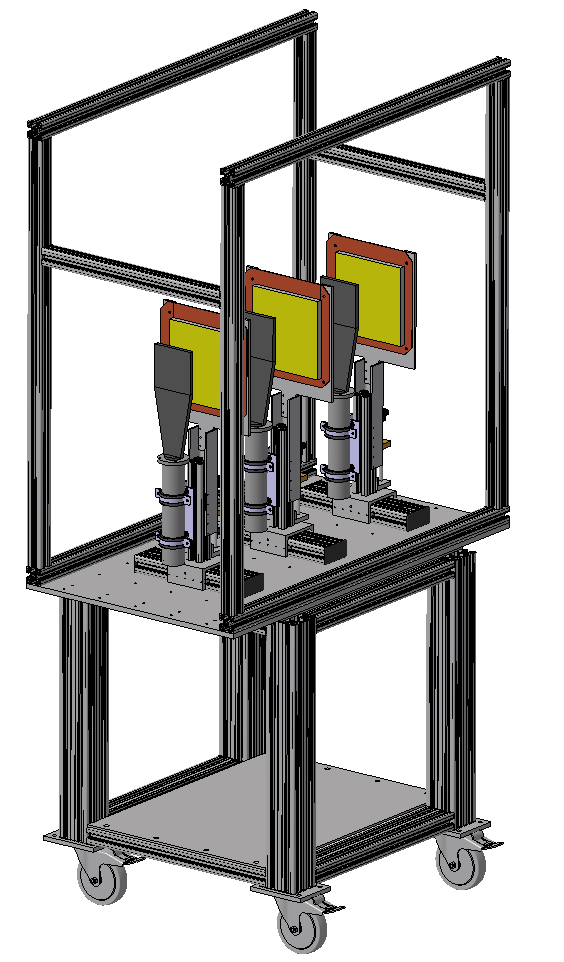}  
        \includegraphics[width=3.8cm, height=5.5cm]{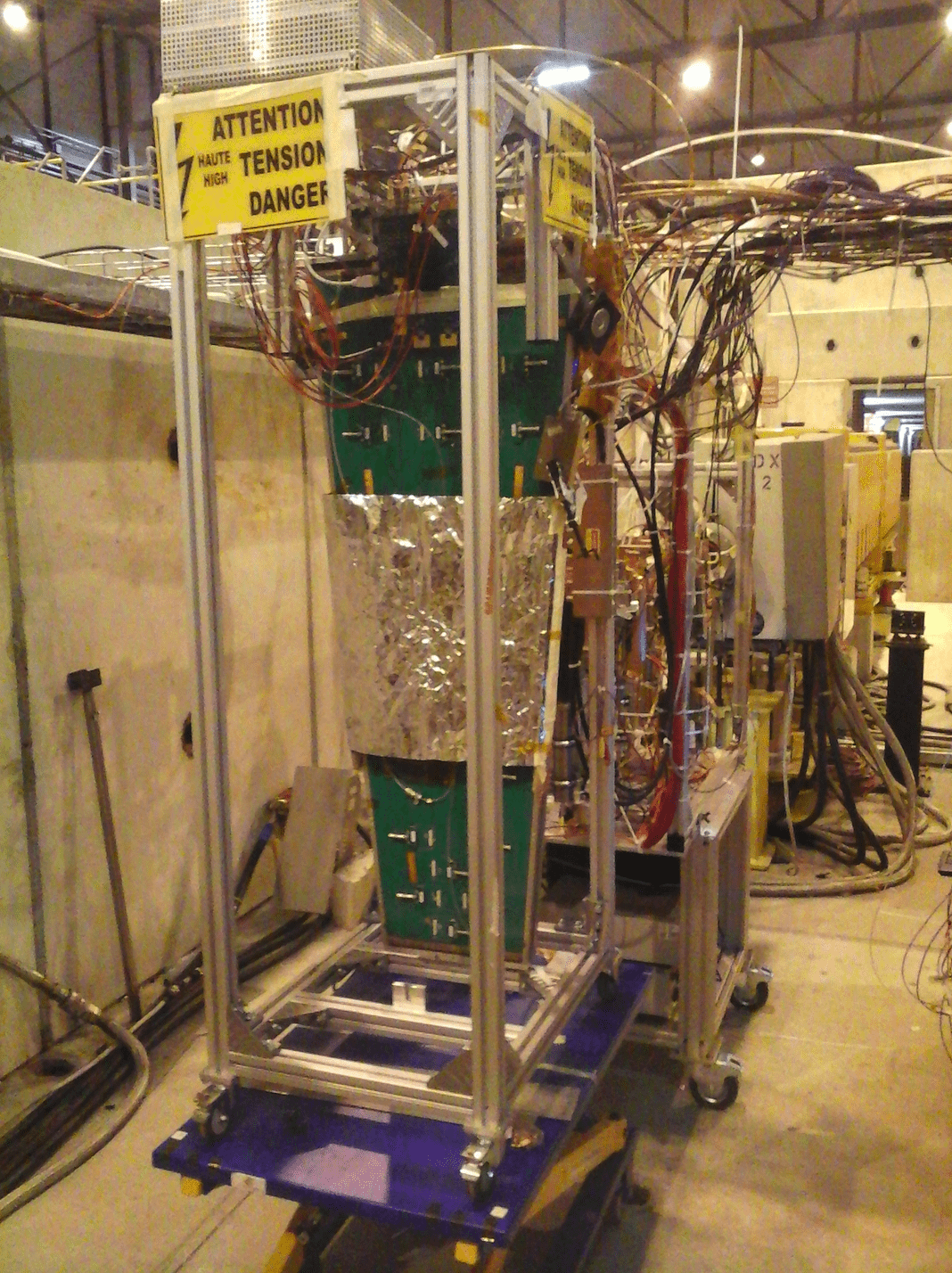} 
        \includegraphics[width=7.5cm, height= 5.5cm]{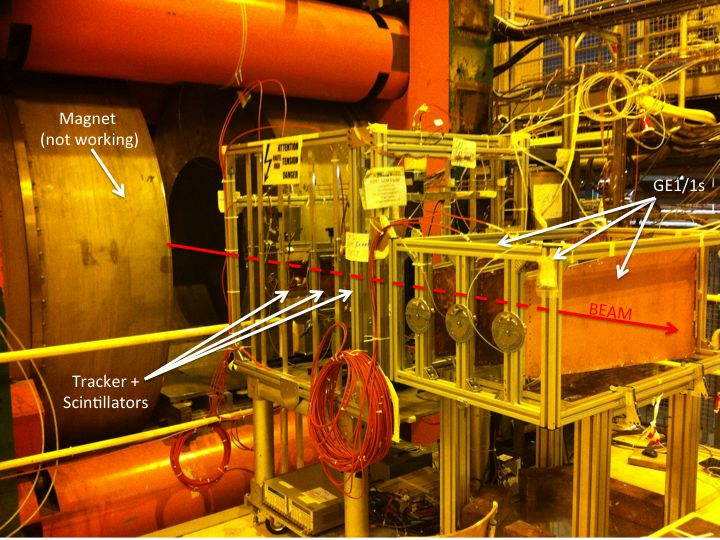}
   \caption{(top-left) Schematic of the beam test setup showing the direction of the muon beam, triggering scintillators (photo multiplier tubes (PMTs)) (in dark gray color), 10 cm $\times$ 10 cm tracking GEMs (in yellow), and GE1/1 chambers (in green)~\cite{five00}, (top-right) design of the tracking telescope showing three triggering scintillators (in grey color), 10 cm $\times$ 10 cm tracking GEMs (in yellow color), (bottom-left) movable aluminum stand holding GE1$\slash$1 chambers in front of the tracking telescope during the H4 beam test campaign, and (bottom-right) one of the earliest (December 2014) beam test setups at the CERN SPS.} \label{fig:Actual_Beam_Setup}
\end{figure}

\noindent 
The test setup, shown in Figure~\ref{fig:Actual_Beam_Setup}, consists of a scintillator hodoscope for triggering, a GEM tracking system to reconstruct the muon trajectory, and the GE1/1 chamber under test.  A movable aluminum support  structure enables  translations in $\phi$ and $\eta$ directions to allow beam alignment with different GE1/1 readout sectors. The test beam setups used in earlier test campaigns can be found in~\cite{three, four, five}. 

\noindent The trigger counters S1, S2, and S3 are organic plastic scintillators with discriminated outputs. A trigger is formed from their triple-coincidence. 

\noindent The tracking telescope, developed by the RD51 Collaboration~\cite{three}, consists of 
three 10 cm $\times$ 10 cm GEM detectors with strip-based two-dimensional readout planes.
The GE1$\slash$1 chambers are aligned perpendicular to the direction
of the muon beam and placed downstream of the tracking telescope. The
chambers are shielded with aluminum and copper clad foils to reduce
the noise level to below that of the expected signals. 


\subsection{Readout Electronics}

\noindent The GE1\slash1 chamber has 3072 readout strips which are organised in groups of 128 to form 8 $\times$ 3  ($\eta$, $\phi$) sectors. The signals are read out by the Application Specific Integrated Circuit (ASIC) VFAT2~\cite{VFAT201, VFAT202}, a digital front end chip (Figure~\ref{fig:Efficiency} (top)) originally developed by the TOTEM Collaboration~\cite{totemCollab}. The chip has 128 readout channels, each connected to a pre-amplifier, shaper, and constant fraction discriminator. The VFAT2 chip provides a binary output with a variable latency for the position information and a fixed latency output, called the SBIT, for time information. 

\noindent The VFAT2 is controlled by TURBO cards, a stand-alone portable control and data acquisition platform developed for TOTEM Test Platforms~\cite{ttp} for the VFAT2. Each TURBO card can accommodate up to 8 VFAT2 chips. The trigger signal is sent to a master TURBO card. The output from the master TURBO card acts as an input to the slave TURBO card. Both cards receive GEM signals from the VFAT2 chips connected to the tracker and GE1/1 chambers. The TURBO boards are controlled through Labview. 

\noindent The output from ($\eta$, $\phi$) sector (5, 2) is sent to
the shaper of the VFAT2 and then compared to a customizable threshold
that was set to 1.2 fC, selected to optimize the process of data acquisition in noisy environments. 

\subsection{Efficiency}
\noindent Efficiency, which represents the probability to record an event when a particle crosses the detector, is estimated by recording the total number of triggers $N$ generated by the coincidence of the three scintillators and the number of hits $N_{1}$, generated by a test region. However, due to possible misalignment of the test region and particle scattering, the number of hits $N_{2}$ are also observed from neighboring regions. Therefore, the efficiency is calculated by removing these additional hits from the total number of triggers using the formula $\varepsilon = N_{1}/(N-N_{2}).$

\begin{figure}[!ht]
    \centering
        \includegraphics[width=3.8cm, height=5cm]{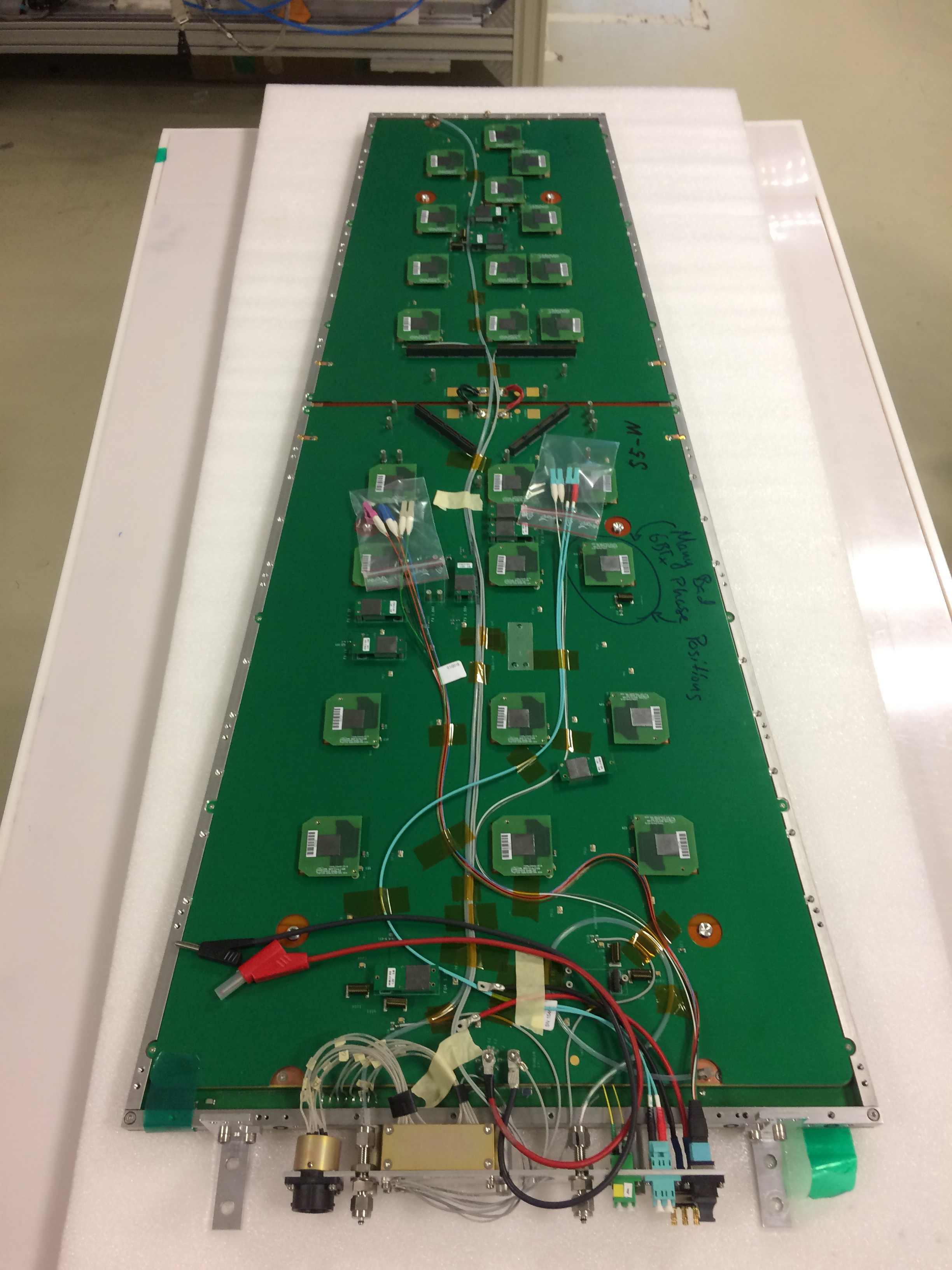}
        \includegraphics[width=7.5cm, height=5.8cm]{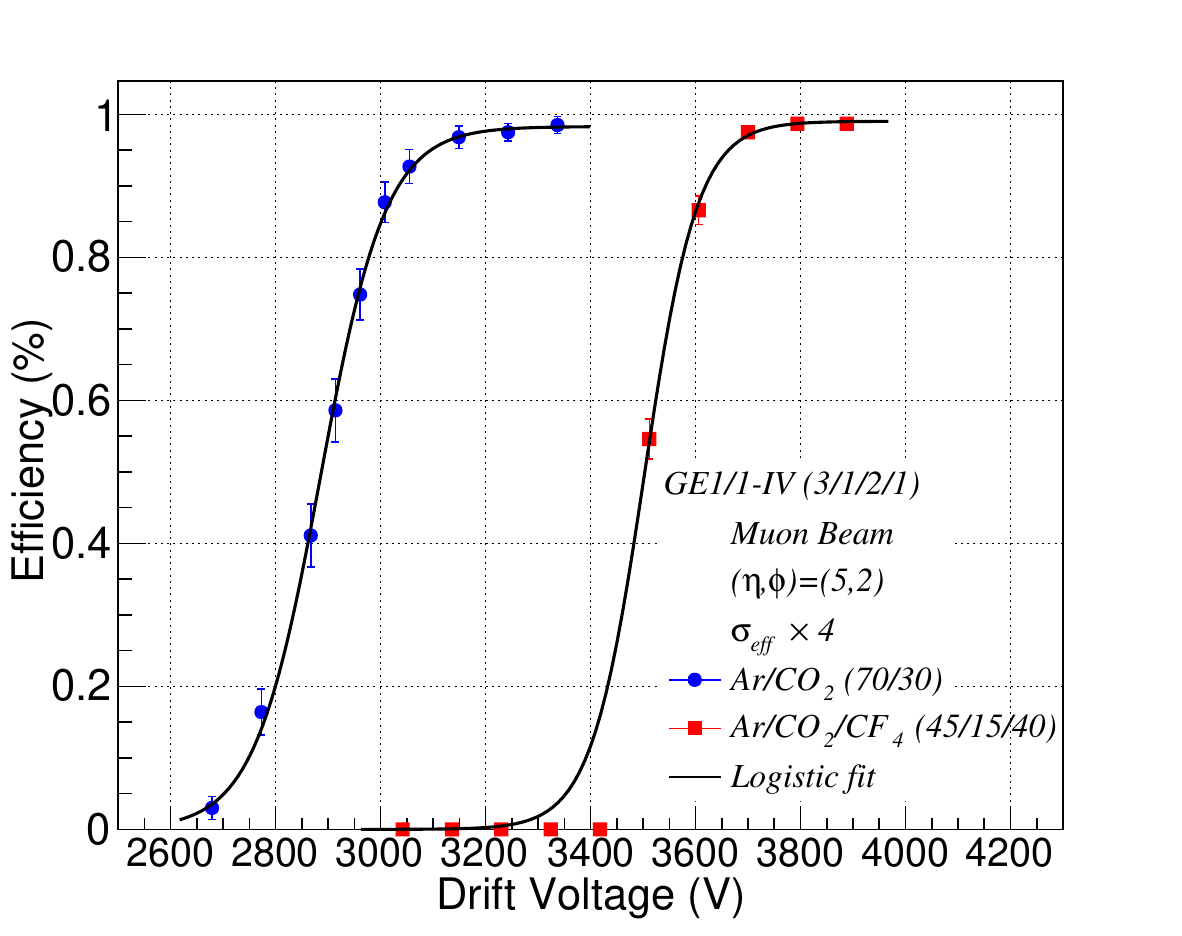}
        \includegraphics[width=7.5cm, height=5.8cm]{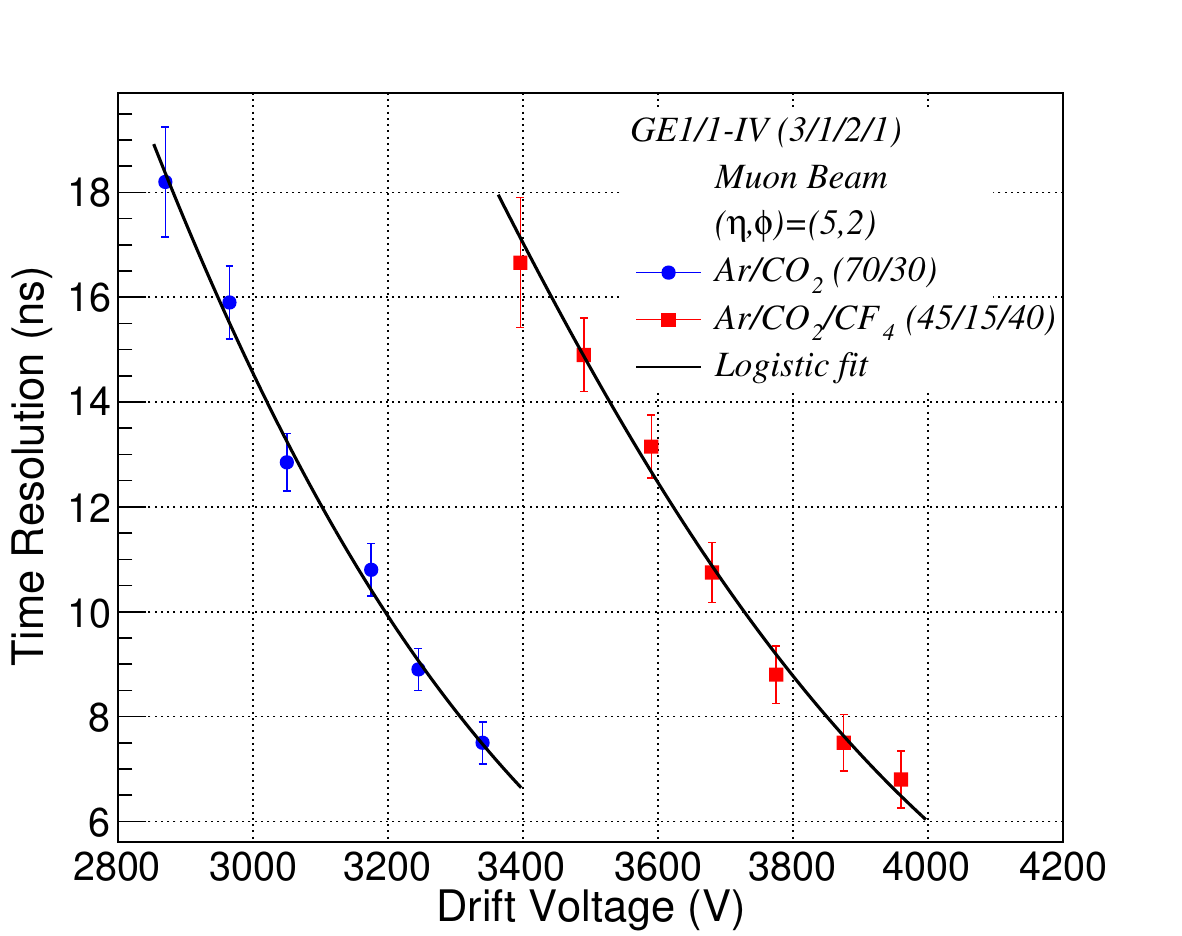} 
   \caption{(top-left) GE1$\slash$1 chamber mounted with 24 VFAT chips, (top-right) efficiency, and (bottom) time resolution of a GE1/1-IV detector for the gas compositions Ar$\slash$CO$_{2}$ (70$\slash$30) and Ar$\slash$CO$_{2}$$\slash$CF$_{4}$ (45$\slash$15$\slash$40). Points represent the data and solid lines represent parameterized fits (details in Table~\ref{table:01}).
The error bars on data represent Gaussian one sigma uncertainty. Since the uncertainty is small, for display, the errors are multiplied by a factor of 4 ($\sigma_{eff}$ $\times$ 4). } \label{fig:Efficiency}
\end{figure}

\noindent The efficiency of the GE1$\slash$1 chambers is measured for the gas mixtures Ar$\slash$CO$_{2}$ (70$\slash$30) and Ar$\slash$CO$_{2}$$\slash$CF$_{4}$ (45$\slash$15$\slash$40) as a function of drift bias. An average efficiency plateau of over 98$\%$ is reached at lower voltages for the Ar$\slash$CO$_{2}$ gas mixture compared to Ar$\slash$CO$_{2}$$\slash$CF$_{4}$ corresponding to effective gains of $\sim$$10^4$  as shown in Figure~\ref{fig:Efficiency}. The drift bias to reach the efficiency plateau is lower for the gas mixture without CF$_{4}$  due to the quenching effect of CF$_{4}$, as discussed in section~\ref{Sect:Results}.

\subsection{Time Resolution}

\noindent Time resolution is an essential parameter because the GE1$\slash$1 detectors, used both in the CMS trigger and off-line reconstruction, must identify the correct bunch crossing. 
Several factors determine the time resolution. There are unavoidable fluctuations in the position of the primary ionization cluster formed in the drift region in which the electric field strength determines the drift velocity. There is charge diffusion in the gaps between the GEM foils which can be reduced by the addition of a gas component with low diffusion coefficient such as CF$_{4}$.  The time resolution is measured by the standard deviation of the distribution of the time difference between the trigger and detector signal. The trigger coincidence signal is sent to the common stop input of a CAEN model V775 TDC and the latency output of the detector is sent to a TDC input. 

\begin{figure}[!ht]
    \centering
        \includegraphics[width=7.5cm, height=6.2cm]{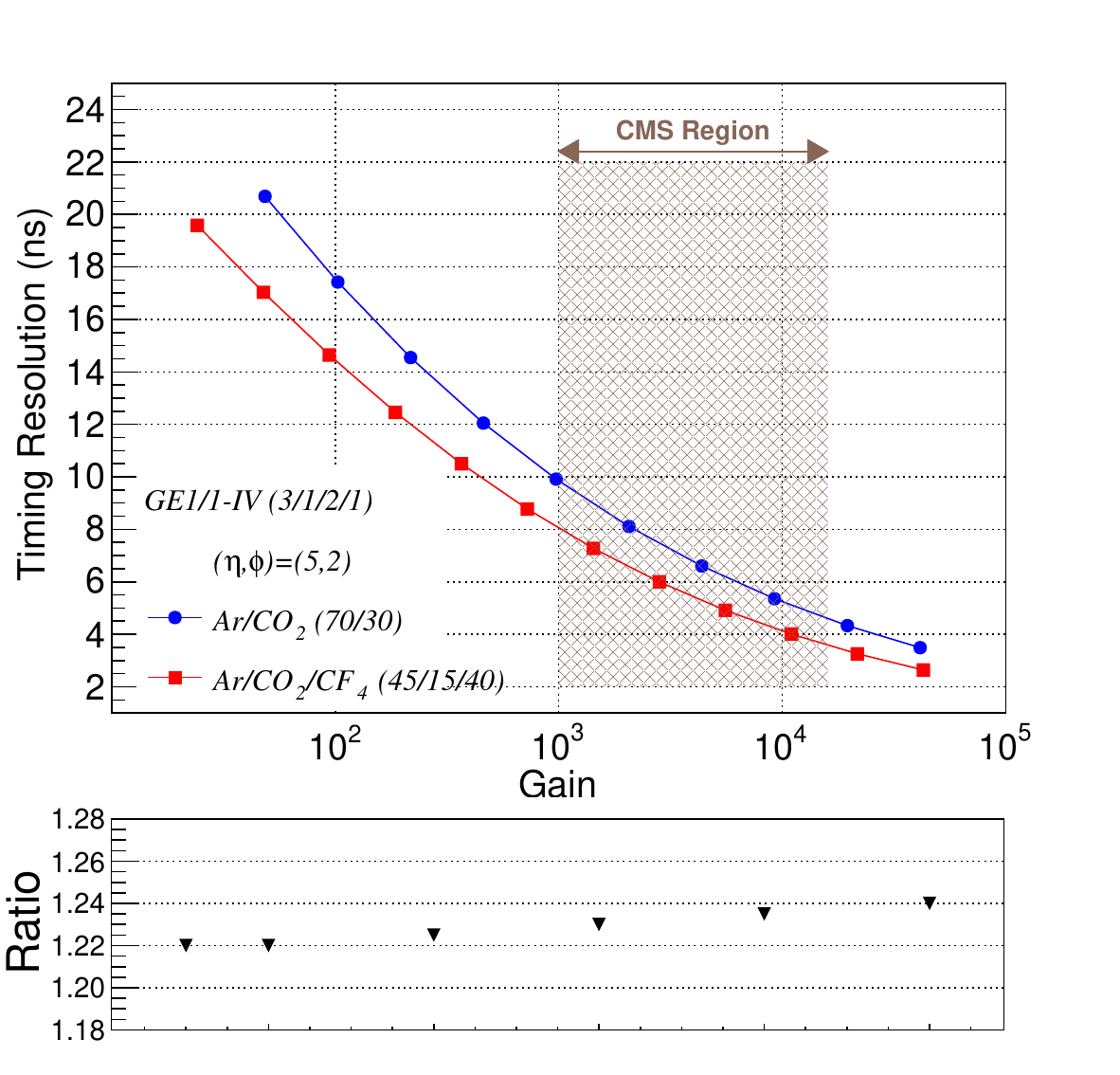}
        \includegraphics[width=7.8cm, height=6.2 cm]{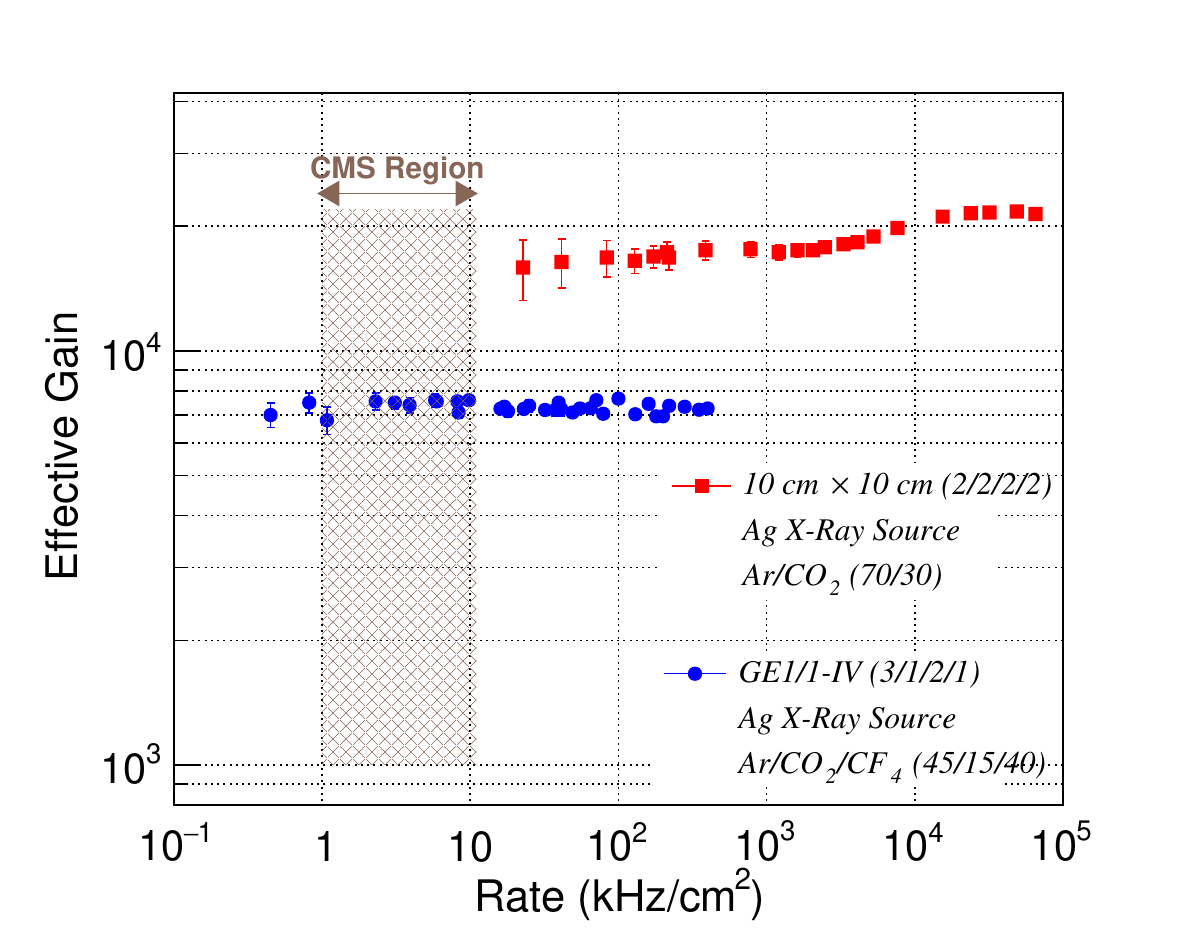}
   \caption{(top) Time resolution for Ar$\slash$CO$_{2}$ (70$\slash$30) and Ar$\slash$CO$_{2}$$\slash$CF$_{4}$ gases as a function of gain. The fit equations from Figure~\ref{fig:Efficiency} (bottom) are used to obtain data points by interpolation; the solid lines connect the points. The ratio of time resolution for Ar$\slash$CO$_{2}$$\slash$CF$_{4}$ (45$\slash$15$\slash$40) to Ar$\slash$CO$_{2}$ (70$\slash$30) is shown in same plot.  (bottom) Rate capabilities of a GE1$\slash$1-IV chamber and, for comparison, a 10 cm $\times$ 10 cm test detector. The shaded ``\texttt{CMS Region}" spans the range of particle flux expected in CMS for HL-LHC.} \label{fig:timing}
\end{figure}

\noindent Measurements are shown in Figure~\ref{fig:Efficiency} for a GE1$\slash$1-IV detector with gas compositions of Ar$\slash$CO$_{2}$ and Ar$\slash$CO$_{2}$$\slash$CF$_{4}$. The time resolution primarily depends upon the drift voltage at a constant value of transfer and induction fields. Because the chambers were powered using a divider chain~\cite{GEMlayout}, the relative strengths of all of the fields were fixed. Therefore, to compare the time resolution achieveable between the two gas mixtures, the measured time  resolution is determined as a function of gain, shown in Figure~\ref{fig:timing}, using fits of time resolution versus drift voltage, shown in Figure~\ref{fig:Efficiency} (bottom). As seen in Figure~\ref{fig:timing}, there is an improvement of $\sim24\%$ by adding a CF$_{4}$ component to Ar$\slash$CO$_{2}$. The \texttt{CMS Region} shows the time resolution for the operating conditions expected in CMS.

\section{Rate Capability} \label{Sect:RateCapability}

\noindent The flux in the CMS end-caps is not expected to exceed 10
kHz/cm$^2$ and the nominal operating gain of a GE1$\slash$1 detector
is expected to be $\sim$ 7 $\times 10^3$ ~\cite{cmsTdr}.  Using an
intense source of X-ray photons, the rate capability is assessed by
measuring the gain as a function of rate. The amplified current is
measured using a pico-ammeter connected to the anode of the detector
as the incident particle flux is varied using copper attenuators. The
measurements (Figure~\ref{fig:timing} (bottom)) show that the effective
gain of a
full-size GE1$\slash$1-IV  detector
remains stable up to several hundreds of kHZ/cm$^2$. Measurements are
also shown for a 10 cm $\times$ 10 cm test detector with
$2\slash2\slash2\slash2$ mm gap configuration at a starting gain of
$\sim$1.5 $\times$ 10$^4$ in Ar/CO$_2$ (70/30)~\cite{SingleMask01,
five03}. The effective gain remains stable up to 10$^5$ kHz/cm$^2$.

\section {Discharge Probability}\label{Sect:DischargeProbability}

\noindent The GE1$\slash$1 detectors will operate at sufficiently high gain ($\sim$$10^4$) to ensure maximum detection efficiency while maintaining timing performance. However, with high gains, intense particle fluxes, and densely ionizing particles, the probability increases for producing discharges that could damage the detectors. Discharges are initiated when local charge exceeds the Raether limit~\cite{rathaerL}, resulting in variations in the local electric field. These variations can transform the avalanche into a streamer that propagates in both directions toward GEM electrodes and provokes electrical breakdown of the gas. However, several features of the chamber design reduce the probability of a discharge and limit the damage from those that occur, as previously established~\cite{six}. These features, discussed below, are the asymmetric distribution of charge-amplifying electric fields over the three GEM foils, sectorization of the GEM foils, and use of protection resistors to limit the available energy in case of a discharge, 

\noindent The three amplification stages of a GE1$\slash$1 chamber are set to slightly different gains by applying different voltages across the foils. The voltage across the first GEM foil is 3$\%$ higher than the second GEM foil, which itself is 5$\%$ higher than the third foil. This configuration significantly reduces the probability of discharge because of the gain sharing between the three amplification stages. Furthermore, because of separate  amplification stages and an independent readout plane, the propagation of a streamer before further amplification is considerably diminished and the probability is reduced for inducing large signals on the readout board, preventing damage to the detector and electronics. 

\noindent The GEM foils are designed to reduce damage from discharges. The electrodes facing the drift plane are divided into several sectors, each with an area ~100 cm$^2$ and each sector with a 10 M$\Omega$ protection resistor. In case of a discharge, the current flowing through the resistor will induce a voltage drop across it, limiting the energy of a discharge and reducing its propagation. 

\begin{figure}[htbp]
    \centering
          \includegraphics[width=7.8cm, height=6.3cm] {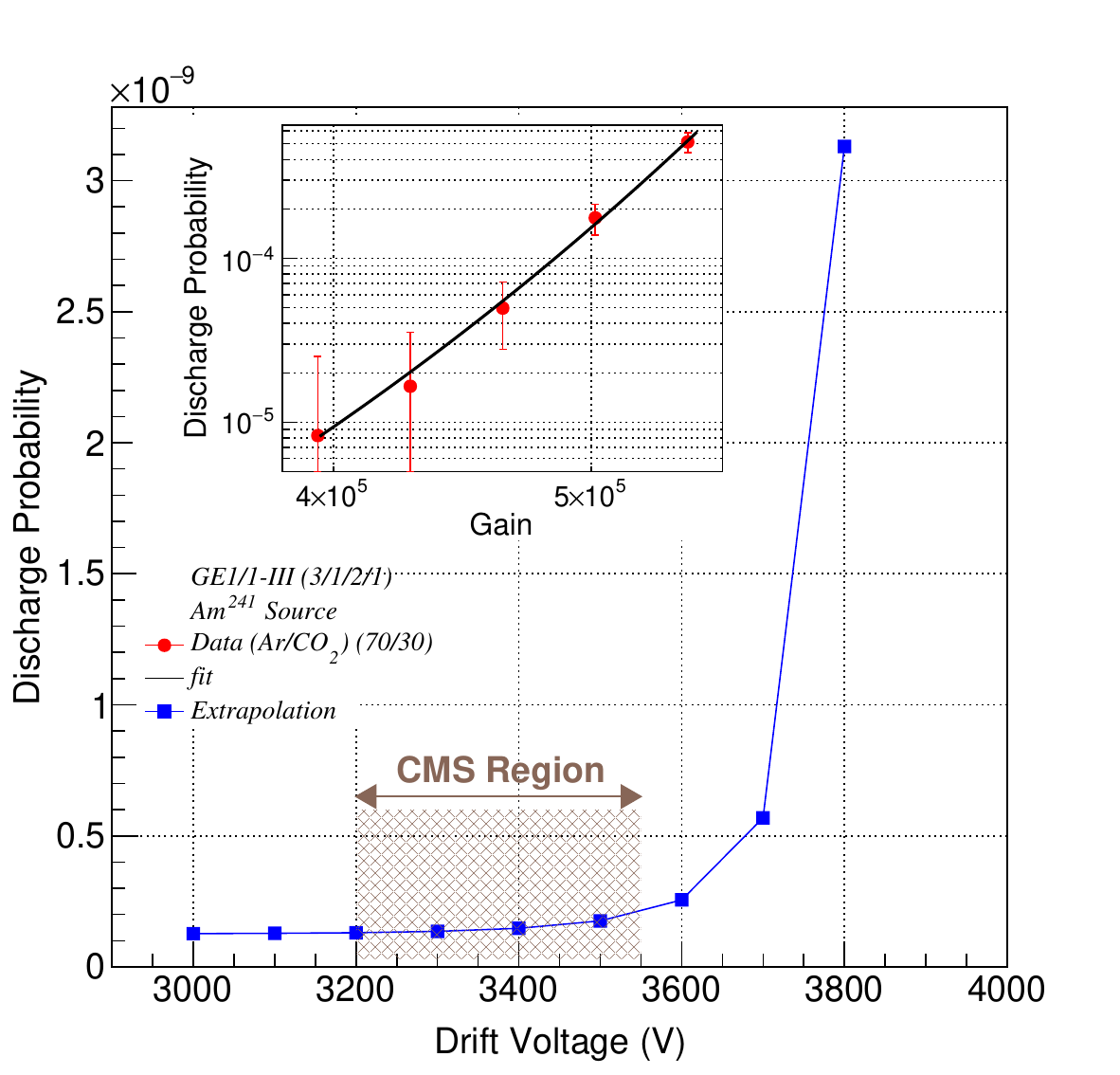} 
   \caption{Discharge probability for the gas composition Ar$\slash$CO$_{2}$.} \label{fig:Discharge_probability}
\vspace{-10pt}
\end{figure}

\noindent The discharge probability is measured for a  third generation GE1$\slash1$~\cite{cmsTdr} detector and, separately, for a 10 cm $\times$ 10 cm test detector with the same gap configuration as a GE1$\slash$1 detector \cite{five03}. In each case, gain is set to extremely high values ranging from 4 to 6 $\times$ $10^5$. The detectors are irradiated by densely ionizing $\alpha$-particles from  a $^{241}$Am source. The measured discharge probability versus gain for a GE1$\slash$1-III detector is shown in Figure~\ref{fig:Discharge_probability}.  The data are also displayed as discharge probability versus drift potential, with the \texttt{CMS Region} indicated. Because an alpha particle from $^{241}$Am produces nearly a hundred times more primaries than a MIP, the discharge probability for a MIP must be divided by this factor.  In the \texttt{CMS Region} this probability is less than $\sim$10$^{-12}$, well within the requirements for the GE1/1 system.

\section {Fits to Data and Summary}

\begin{figure}[!ht]
    \centering
        \includegraphics[width=9cm,height=7.5cm] {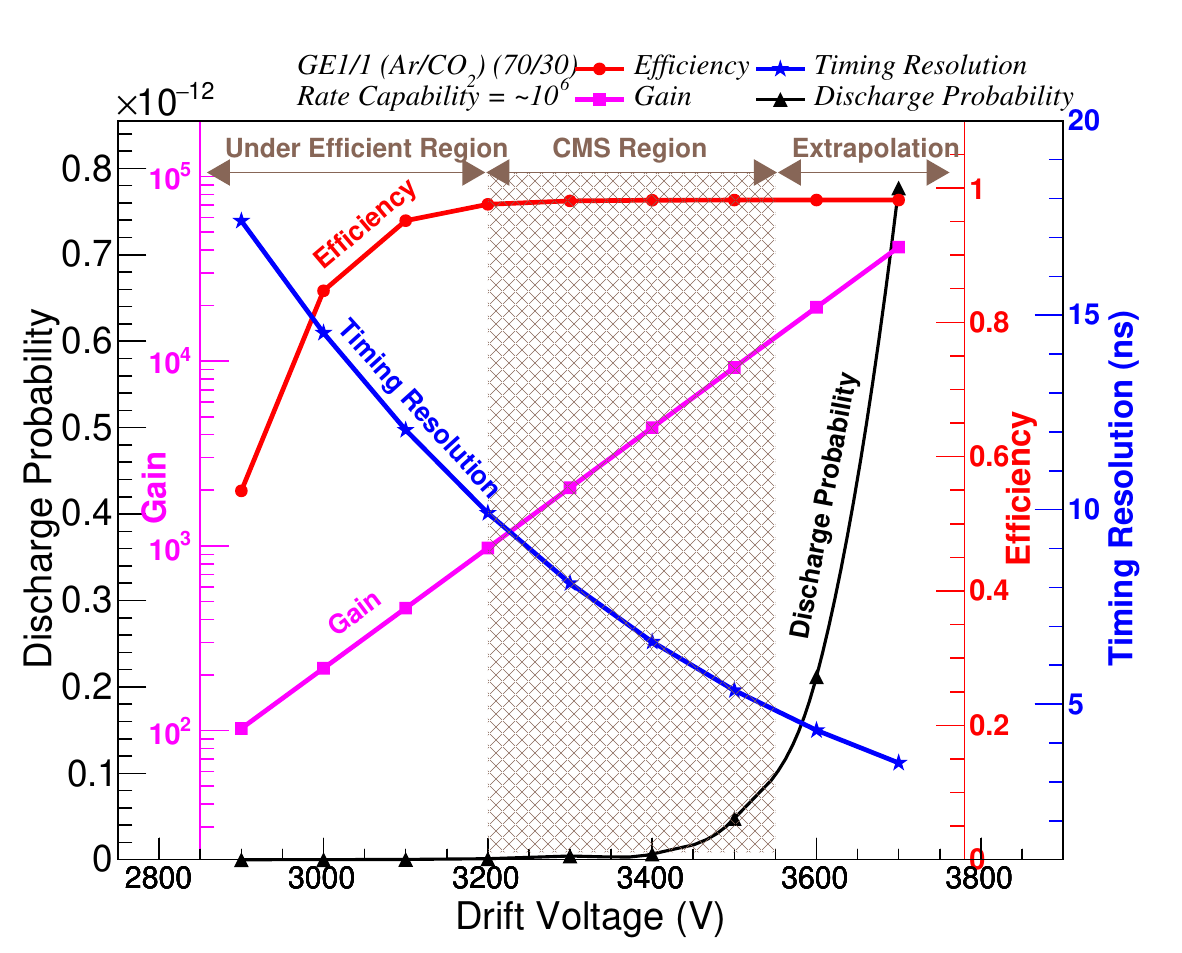} 
   \caption{Master plot of GE1$\slash$1 detectors showing the gain (pink), discharge probability (black), efficiency (red), and time resolution (blue) for the gas composition Ar$\slash$CO$_{2}$ (70$\slash$30) as a function of drift voltage. The axes and corresponding data are represented by the unique color code in the plot. Also, the plot shows the shaded region that is the recommended operational region of the chambers during their use in CMS.} \label{fig:master_plot_ArCO2}
\vspace{-10pt}

\end{figure}

\begin{figure}[!ht]
    \centering
        \includegraphics[width=9cm,height=7.5cm]{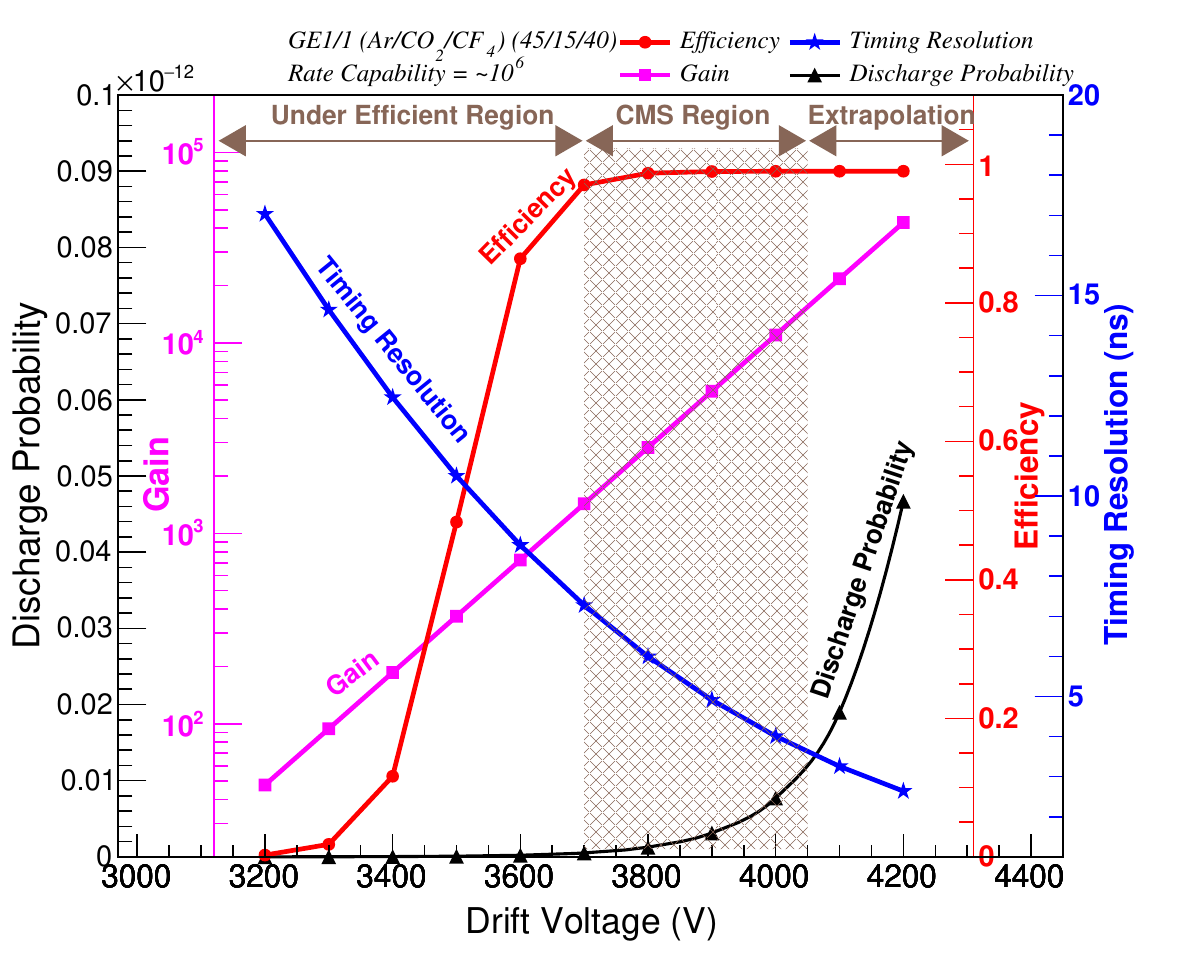} 
   \caption{Master plot of GE1$\slash$1 detectors showing the gain (pink), discharge probability (black), efficiency (red), and time resolution (blue) for the gas composition Ar$\slash$CO$_{2}$/CF$_{4}$ (45$\slash$15$\slash$40) as a function of drift voltage.  The axes and corresponding data are represented by the unique color code in the plot. Also, the plot shows the shaded region that is the recommended operational region of the chambers during their use in CMS.} \label{fig:master_plot_ArCO2CF4}
\end{figure}

To characterize the performance of a GE1$\slash$1 chamber for any drift voltage, the data for gain, discharge probability, efficiency, and time resolution are fit with parametric equations as shown in Table~\ref{table:01}. The fits provide a good description of the data and allow interpolation to any desired value of drift voltage. Interpolated data points are obtained for the  measured quantities and are displayed in the master plots of Figures \ref{fig:master_plot_ArCO2} and \ref{fig:master_plot_ArCO2CF4} for  Ar$\slash$CO$_{2}$ and Ar$\slash$CO$_{2}\slash$CF$_{4}$ gases, respectively.  The \texttt{CMS Region}, shown as a shaded area in the figures, corresponds to the operational regime for the CMS experiment.

\begin{table}[!ht]
\centering
\tiny
\begin{tabular}{l|l|l|l|l|l} 
\hline
\hline
Gas 	    &Property    &a    &b    &c    &Fit Equation \\        
\hline
\hline
Ar/CO$_{2}$        &Gain    &270.85    &-22.72    &0.007   & $G  = a e^{(b+cV)}$\\
\hline

    &Efficiency    &0.983    &2885.12    &62.59    & $\varepsilon = a / (1+e^{-(V-b)/c} )$\\
\hline
        &Time Resolution    &78.48    &2346.92    &441.1    & $R = a / (1+e^{(V-b)/c} )$\\
\hline
        &Discharge Probability    & $1.002 \times 10^{-31}$    &0    &0.013    & $DP = a e^{(b+cV)}$\\
\hline
\hline
Ar/CO$_{2}$/CF$_{4}$        &Gain    &4336.03    &-26.27    &0.0068    &$G = a e^{(b+cV)}$\\
\hline

    &Efficiency    &0.99    &3502.37    &50.78    &$\varepsilon = a / (1+e^{-(V-b)/c})$\\
\hline
        &Time Resolution   &39.21    &3126.77    &428.09    &$R = a / (1+e^{(V-b)/c}$)\\
\hline
        &Discharge Probability    & $1.79 \times 10^{-28}$    &0    &0.009    & $DP = a e^{(b+cV)}$\\
\hline
\hline
\end{tabular}
\caption{Equations and parameter values from fits to data for gain, efficiency, time resolution, and discharge probability versus drift voltage for gas compositions Ar$\slash$CO$_{2}$ and Ar$\slash$CO$_{2}\slash$CF$_{4}$.}
\label{table:01}
\vspace{-10pt}
\end{table}
 
In summary, different generations of GE1$\slash$1 detectors are tested for gain, efficiency, time resolution, and discharge probability with gas compositions of Ar$\slash$CO$_{2}$ and Ar$\slash$CO$_{2}\slash$CF$_{4}$. The measurements show that  a GE1$\slash$1 detector
can be operated up to a gain of about $10^5$ with a discharge
probability of less than $10^{-11}$ under CMS operating conditions
with a MIP rate up to 10$^{6}$ Hz. The performance of the chamber in beam tests shows an efficiency of 98$\%$ or better obtained across the active area, and time resolution close to 5 ns.



\vspace{50pt}

\section*{Acknowledgements}
We gratefully acknowledge the support from FRS-FNRS (Belgium), FWO-Flanders (Belgium), BSF-MES (Bulgaria), BMBF (Germany), CSIR \& UGC (India), DAE (India), DST (India), INFN (Italy), NRF (Korea), LAS (Lithuania), QNRF (Qatar), DOE (USA) and the RD51 collaboration.

\end{document}